\documentclass[aps,prb,twocolumn]{revtex4}

\usepackage{graphicx}
\usepackage{bm}

\begin{document}

\title{N\'eel order in the Hubbard model within spin-charge rotating\\
 reference frame approach: crossover from weak to strong coupling}

\author{T. A. Zaleski and T. K. Kope\'{c}}

\affiliation{Institute for Low Temperature and Structure Research,\\
 Polish Academy of Sciences,\\
 POB 1410, 50-950 Wroclaw 2, Poland}

\date{\today}

\begin{abstract}
The antiferromagnetic phase of two-dimensional ($2D$) and three-dimensional
($3D$) Hubbard model with nearest neighbors hopping is studied on
a bipartite cubic lattice by means of the quantum SU(2)$\times$U(1)
rotor approach that yields a fully self-consistent treatment of the
antiferromagnetic state that respects the symmetry properties the
model and satisfy the Mermin-Wagner theorem. The collective variables
for charge and spin are isolated in the form of the space-time fluctuating
U(1) phase field and rotating spin quantization axis governed by the
SU(2) symmetry, respectively. As a result interacting electrons appear
as a composite objects consisting of bare fermions with attached U(1)
and SU(2) gauge fields. An effective action consisting of a spin-charge
rotor and a fermionic fields is derived as a function of the Coulomb
repulsion $U$ and hopping parameter $t$. At zero temperature, our
theory describes the evolution from a Slater $(U\ll t)$ to a Mott-Heisenberg
$(U\gg t)$ antiferromagnet. The results for zero-temperature sublatice
magnetization ($2D$) and finite temperature ($3D$) phase diagram
of the antiferromagnetic Hubbard model as a function of the crossover
parameter $U/t$ are presented and the role of the spin Berry phase
in the interaction driven crossover is analyzed. 
\end{abstract}


\maketitle

\section{Introduction}

Effective low-energy theories are frequently employed in essentially
all fields of physics. In the realm of strongly correlated electrons,
spin only Hamiltonians, are examples of effective low-energy theories
that apply in the limit of strong interactions. Due to the numerical
expense in solving these models for large lattices, it is imperative
to look for methods which relay on advanced analytical approaches.
A key question in this context concerns the emergence of low energy
scales, much smaller than the bare Coulomb interactions between the
electrons, which govern the existence and the competition of different
phases. This can be studied by considering prototypical lattice models
of strongly correlated electrons. In the strong coupling limit, the
$t-J$ Hamiltonian,\cite{spalek} derived from the large$-U$ Hubbard
model,\cite{hubbard} is often used to describe the low lying excitations.
At half-filling, it directly reduces to the quantum antiferromagnetic
Heisenberg model. However, for finite $U$ as in the the Hubbard model,
the quantum-mechanical objects are not local spins since one has still
mobile electrons and one expects that calculation of the ground-state
phase diagram as a function of the density and interaction strength
is more difficult than for the Heisenberg model. In the weak coupling
limit $(U\ll t)$, a Fermi surface instability gives rise to a spin-density-wave
ground state as described by Slater,\cite{slater} where the antiferromagnetic
(AF) long-range order produces a gap in the quasiparticle excitation
spectrum. In the strong-coupling regime $(U\gg t)$, fermions are
effected by the strong Coulomb repulsion causing the Mott--Hubbard
localization.\cite{mott} The created local magnetic moments subsequently
order at low temperature and give rise to a Mott-Heisenberg antiferromagnet.
It is well known that even for small but finite interactions the Hartree-Fock
(HF) N\'eel temperature is proportional to $U$, which is unrealistically
high since $U$ is a large energy scale, of the order of electron
volts. This wrong prediction had to be expected since correlations
are absent in the HF approach. The requirement of self-consistency
by incorporating spin and charge fluctuations, while maintaining the
essential spin-rotation symmetry, summarizes the challenging nature
of magnetic ordering in strongly correlated systems. A variety of
theoretical methods are available for the study of strongly correlated
systems. In weak or strong coupling perturbative treatments usually
are used.\cite{imada} On the other hand numerical -- very effective
in low dimensional models -- are limited to finite systems, which
requires an extrapolation to the thermodynamic limit that is often
problematic. As a result the numerical approach does not provide in
general a unifying picture, which only analytical approaches can
give.

The objective of the present paper is to quantitatively investigate
correlation effects in a antiferromagnetic state of the Hubbard model
within a spin-rotationally-symmetric scheme that is fully compatible
with the Mermin-Wagner theorem.\cite{mermin} To this end, we describe
a theoretical approach which provides a unified view of two $(2D)$
and three $(3D)$ dimensional model at half-filled Hubbard model for
any value of the Coulomb repulsion $U$, which is able to handle the
evolution from the Slater to the Mott-Heisenberg antiferromagnet that
captures correctly both the spin and charge degrees of freedom. We
address the above questions by implementing the charge-U(1) and spin-SU(2)
rotationally invariant handling of the Hubbard model. By recognizing
spin and charge symmetries we explicitly factorize the charge and
spin contribution to the original electron operator in terms of the
corresponding gauge fields that leads to a composite particle, which
is the union of an electron with U(1) and SU(2) gauge potentials.
In this scheme the charge and spin excitations emerge in terms of
a U(1) phase and variable spin quantization axis: the effective field
theory for the strongly correlated problem is thus chracterized by
the U(2)=U(1)$\times$SU(2) group, where the gauge potential in U(1)
describes the evolution of a particle scalar characteristic, which
is naturally associated with an electric charge, while the gauge potential
in SU(2) describes the nontrivial dynamics associated with the evolution
of the vector internal characteristic of a particle such as spin.

The outline of the paper is as follows: In Secs. II we introduce the
model and, in Sec.III-V, we develop the analytical background needed
for the calculations. In Section VI we find a closed set of self-consistent
equations for the antiferromagnetic gap and order parameter, while
the numerical evaluation of self-consistent equations is presented
in Secs. VII, where the phase diagrams considering the paramagnetic
the antiferromagnetic phase for the Hubbard model and in different
dimensionality are calculated. 

\section{The model}

Our starting point is the purely fermionic Hubbard Hamiltonian ${\cal H}\equiv{\cal H}_{t}+{\cal H}_{U}$:
\begin{eqnarray}
{\cal H}=-t\sum_{\langle{\bf r}{\bf r}'\rangle,\alpha}[c_{{\alpha}}^{\dagger}({\bf r})c_{\alpha}({\bf r}')+{\rm h.c.}]+U\sum_{{\bf r}}n_{\uparrow}({\bf r})n_{\downarrow}({\bf r}).\label{mainham}\end{eqnarray}
 Here, $\langle{\bf r},{\bf r}'\rangle$ runs over the nearest-neighbor
(n.n.) sites, $t$ is the hopping amplitude, $U$ stands for the Coulomb
repulsion, while the operator $c_{\alpha}^{\dagger}({\bf r})$ creates
an electron with spin $\alpha=\uparrow,\downarrow$ at the lattice
site ${\bf r}$, where ${n}_{\alpha}({\bf r})=c_{\alpha}^{\dagger}({\bf r})c_{\alpha}({\bf r})$.
Usually, working in the grand canonical ensemble a term is added to
${\cal H}$ in Eq. (\ref{mainham}) to control the average number
of electrons, ${\cal H}\to{\cal H}-\mu\sum_{{\bf r}}{n}({\bf r})$
with $\mu$ being the chemical potential and ${n}({\bf r})=n_{\uparrow}({\bf r})+n_{\downarrow}({\bf r})$
the fermionic number operator. 

\subsection{Grassman action}

The functional integral representation of models for correlated electrons
allows us to implement efficiently the method of treatment. It is
customary to introduce Grassmann fields, $c_{\alpha}({\bf r}\tau)$
depending on the {}``imaginary time\char`\"{} $0\le\tau\le\beta\equiv1/k_{B}T$,
(with $T$ being the temperature) that satisfy the anti--periodic
condition $c_{\alpha}({\bf r}\tau)=-c_{\alpha}({\bf r}\tau+\beta)$,
to write the path integral for the statistical sum ${\cal Z}=\int\left[{\cal D}\bar{c}{\cal D}{c}\right]e^{-{\cal S}[\bar{c},c]}$
with the fermionic action \begin{eqnarray}
{\cal S}[\bar{c},c]={\cal S}_{B}[\bar{c},c]+\int_{0}^{\beta}d\tau{\cal H}[\bar{c},c]\label{cberry}\end{eqnarray}
 that contains the fermionic Berry term:\cite{berry} \begin{equation}
{\cal S}_{B}[\bar{c},c]=\sum_{{\bf r}\alpha}\int_{0}^{\beta}d\tau\bar{c}_{\alpha}({\bf r}\tau)\partial_{\tau}{c}_{\alpha}({\bf r}\tau),\end{equation}
 which will play an important role in our considerations. 

\section{SU(2)$\times$U(1) action}

It is customary to introduce auxiliary fields for the spin and charge
fluctuations via a Hubbard- Stratonovitch (HS) transformation to decouple
the interaction term in the Hubbard Hamiltonian. However, such a procedure
usually leads to a loss of the spin rotational invariance. For strongly
correlated system in order to properly account for the nature of elementary
excitation it is crucial to construct a formulation of the theory
which naturally preserves the existing symmetry present in the Hubbard
Hamiltonian. For this purpose the density--density product in Eq.
(\ref{mainham}) we write, following Ref. \onlinecite{schulz},
in a spin-rotational invariant way: \begin{equation}
{\cal H}_{U}=U\sum_{{\bf r}}\left\{ \frac{1}{4}{n}^{2}({\bf r}\tau)-\left[{\bf \Omega}({\bf r}\tau)\cdot{\bf S}({\bf r}\tau)\right]^{2}\right\} ,\label{huu}\end{equation}
 where $S^{a}({\bf r}\tau)=\frac{1}{2}\sum_{\alpha\alpha'}c_{\alpha}^{\dagger}({\bf r}\tau)\hat{\sigma}_{\alpha\alpha'}^{a}c_{\alpha'}({\bf r}\tau)$
denotes the vector spin operator ($a=x,y,z$) with $\hat{\sigma}^{a}$
being the Pauli matrices. The unit vector \begin{eqnarray}
{\bf \Omega}({\bf r}\tau) & = & [\sin\vartheta({\bf r}\tau)\cos\varphi({\bf r}\tau),\sin\vartheta({\bf r}\tau)\sin\varphi({\bf r}\tau),\nonumber \\
 &  & \cos\vartheta({\bf r}\tau)]\end{eqnarray}
 written in terms of polar angles labels varying in space-time spin
quantization axis. In order to maintain spin rotational invariance,
one should consider the spin-quantization axis to be a priori arbitrary
and integrate over all possible directions of ${\bf \Omega}({\bf r}\tau)$
in the partition function. The reason for choosing the decomposition
in Eq. (\ref{huu}) is that it allows to recover the Hartree-Fock
solution at the saddle point level within the functional integral
formalism. By decoupling spin and charge density terms in Eq. (\ref{huu})
using auxiliary fields $\varrho({\bf r}\tau)$ and $iV({\bf r}\tau)$,
respectively we write down the partition function in the form \begin{eqnarray}
{\cal Z} & = & \int[{\cal D}{\bf \Omega}]\int[{\cal D}V{\cal D}{\varrho}]\int\left[{\cal D}\bar{c}{\cal D}c\right]\times\nonumber \\
 & \times & e^{-{\cal S}\left[{\bf \Omega},V,{\varrho},\bar{c},c\right]},\label{zfun}\end{eqnarray}
 where $[{\cal D}{\bf \Omega}]\equiv\prod_{{\bf r}\tau_{k}}\frac{\sin\vartheta({\bf r}\tau_{k})d\vartheta({\bf r}\tau_{k})d\varphi({\bf r}\tau_{k})}{4\pi}$
is the spin-angular integration measure. The effective action reads:
\begin{eqnarray}
{\cal S}\left[{\bf \Omega},V,{\varrho},\bar{c},c\right] & = & \sum_{{\bf r}}\int_{0}^{\beta}d\tau\left[\frac{{\varrho}^{2}({\bf r}\tau)}{U}+\frac{V^{2}({\bf r}\tau)}{U}\right.\nonumber \\
 & + & \left.iV({\bf r}\tau)n({\bf r}\tau)+2{\varrho}({\bf r}\tau){\bf \Omega}({\bf r}\tau)\cdot{\bf S}({\bf r}\tau)\right]\nonumber \\
 & + & {\cal S}_{B}[\bar{c},c]+\int_{0}^{\beta}d\tau{\cal H}_{t}[\bar{c},c].\label{sa}\end{eqnarray}
We devise a systematic way of decomposing the fluctuating fields contained
in the action in Eq. (\ref{sa}) that enables us to obtain a low energy
effective theory. In the following introduce bosonic fields describing
charge and spin fluctuations, and the fluctuating spin-quantization
axis. 

\subsection{U(1) rotor charge frame}

We observe now that the Hubbard Hamiltonian has a local U(1) gauge
symmetry, when expressed in terms of the underlying electron variables.
This points out a possibility of an emergent U(1) dynamical gauge
field as a fluctuating complex field attached to fermionic variables,
which is dynamically generated, by interacting fermions. In the modern
language it is called a fermion-flux composite. Technically, the appearance
of the U(1) field is based on the Hubbard-Stratonovich decoupling
of the four-fermion interaction -- a typical way to {}``bosonize''
a fermionic system in higher-dimensional problems. The essence of
the method is to eliminate a mixed fermion-boson term in the resulting
action by a gauge transformation.\cite{kopec} The U(1) formulation
begins by rewriting the electron as the product of a charge-neutral
fermionic spinon and a charge-carrying U(1) rotor, which is constrained
to lie on the unit circle in the complex plane. To this end, we write
the fluctuating {}``imaginary chemical potential\char`\"{} $iV({\bf r}\tau)$
as a sum of a static $V_{0}({\bf r})$ and periodic function $V({\bf r}\tau)=V_{0}({\bf r})+\tilde{V}({\bf r}\tau)$
using Fourier series \begin{eqnarray}
\tilde{V}({\bf r}\tau)=\frac{1}{\beta}\sum_{n=1}^{\infty}[\tilde{V}({\bf r}\omega_{n})e^{i\omega_{n}\tau}+c.c.]\label{decomp}\end{eqnarray}
 with $\omega_{n}=2\pi n/\beta$ ($n=0,\pm1,\pm2$) being the (Bose)
Matsubara frequencies. Now, we introduce the U(1) \textit{phase}  field
${\phi}({\bf r}\tau)$ via the Faraday--type relation \begin{equation}
\dot{\phi}({\bf r}\tau)\equiv\frac{\partial\phi({\bf r}\tau)}{\partial\tau}=e^{-i\phi({\bf r}\tau)}\frac{1}{i}\frac{\partial}{\partial\tau}e^{i\phi({\bf r}\tau)}=\tilde{V}({\bf r}\tau).\label{jos}\end{equation}
 Furthermore, by performing the local gauge transformation to the
\textit{new} fermionic variables $f_{\alpha}({\bf r}\tau)$: \begin{equation}
\left[\begin{array}{c}
c_{\alpha}({\bf r}\tau)\\
\bar{c}_{\alpha}({\bf r}\tau)\end{array}\right]=\left[\begin{array}{cc}
z({\bf r}\tau) & 0\\
0 & \bar{z}({\bf r}\tau)\end{array}\right]\left[\begin{array}{c}
f_{\alpha}({\bf r}\tau)\\
\bar{f}_{\alpha}({\bf r}\tau)\end{array}\right]\label{sing1}\end{equation}
 where the unimodular parameter $|z({\bf r}\tau)|^{2}=1$ satisfies
$z({\bf r}\tau)=e^{i\phi({\bf r}\tau)}$, we remove the imaginary
term $i\int_{0}^{\beta}d\tau\tilde{V}({\bf r}\tau)n({\bf r}\tau)$
for all the Fourier modes of the $V({\bf r}\tau)$ field, except for
the zero frequency. Accordingly, the integration measure over the
group manifold becomes 
\begin{equation}
\int[{{\cal D}{\phi}}]\equiv\sum_{\{m({\bf r})\}}\prod_{{\bf r}}\int_{0}^{2\pi}d\phi_{0}({\bf r})\int\limits _{\phi({\bf r}0)=\phi_{0}({\bf r})}^{\phi({\bf r}\beta)=\phi_{0}({\bf r})+2\pi m({\bf r})}{\cal D}\phi({\bf r}\tau).\label{measure}\end{equation}
Since the homotopy group $\pi_{1}[U(1)]$ forms a set of integers,
discrete configurations of $\phi({\bf r}\tau)$ matter, for which
$\phi({\bf r}\beta)-\phi({\bf r}0)=2\pi m({\bf r})$, where $m({\bf r})=0,\pm1,\pm2,\dots$.
Here, $m\in{Z}$ labels equivalence classes of homotopically connected
paths. Thus the paths can be divided into topologically distinct classes,
characterized by a winding number defined as the net number of times
the world line wraps around the system in the {}``imaginary time''
direction.\cite{schulman} 

\subsection{Rotating SU(2) spin reference frame}

Subsequent SU(2) transformation from $f_{\alpha}({\bf r}\tau)$ to
$h_{\alpha}({\bf r}\tau)$ operators, \begin{eqnarray}
\left[\begin{array}{c}
f_{1}({\bf r}\tau)\\
{f}_{2}({\bf r}\tau)\end{array}\right]^{T}={\bf R}({\bf r}\tau)\left[\begin{array}{c}
h_{1}({\bf r}\tau)\\
{h}_{2}({\bf r}\tau)\end{array}\right]\label{sing2}\end{eqnarray}
 takes away the rotational dependence on ${\bf \Omega}({\bf r}\tau)$
in the spin sector. This parametrization makes clear that the SU(2)
matrix rotor is identical to the more familiar O(4) rotor, a quantum
particle constrained to the 3-sphere 
\begin{eqnarray}
{\bf R}({\bf r}\tau)=\left[\begin{array}{cc}
e^{-\frac{i}{2}(\varphi+\chi)}\cos\left(\frac{\vartheta}{2}\right) & -e^{-\frac{i}{2}(\varphi-\chi)}\sin\left(\frac{\vartheta}{2}\right)\\
e^{\frac{i}{2}(\varphi-\chi)}\sin\left(\frac{\vartheta}{2}\right) & e^{\frac{i}{2}(\varphi+\chi)}\cos\left(\frac{\vartheta}{2}\right)\end{array}\right]\end{eqnarray}
with the Euler angular variables $\varphi({\bf r}\tau),\vartheta({\bf r}\tau)$
and $\chi({\bf r}\tau)$, respectively. The link between SO(3)and
SU(2) rotations is established by means of the Hopf map\cite{wilczek}
\begin{equation}
{\bf R}({\bf r}\tau)\hat{\sigma}^{z}{\bf R}^{\dagger}({\bf r}\tau)=\hat{{\bm\sigma}}\cdot{\bf \Omega}({\bf r}\tau)\end{equation}
 that is based on the enlargement from two-sphere $S_{2}$ to the
three-sphere $S_{3}\sim SU(2)$. Here, the extra variable $\chi({\bf r}\tau)$
represents the U(1) gauge freedom of the theory as a consequence of
$S_{2}\to S_{3}$ mapping. One can summarize Eqs (\ref{sing1}) and
(\ref{sing2}) by the single joint gauge transformation exhibiting
electron operator factorization \begin{eqnarray}
c_{\alpha}({\bf r}\tau)=\sum_{\alpha'}{\mathcal{U}}_{\alpha\alpha'}({\bf r}\tau)h_{\alpha'}({\bf r}\tau),\label{decomp2}\end{eqnarray}
 where \begin{equation}
{\mathcal{U}}({\bf r}\tau)=z({\bf r}\tau){\bf R}({\bf r}\tau)\end{equation}
is a U(2) matrix which rotates the spin-charge quantization axis at
site ${\bf r}$ and time $\tau$. Eq. (\ref{decomp2}) reflects the
composite nature of the interacting electron formed from bosonic spinorial
and charge degrees of freedom given by ${R}_{\alpha\alpha'}({\bf r}\tau)$
and $z({\bf r}\tau)$, respectively as well as remaining fermionic
part $h_{\alpha}({\bf r}\tau)$. In the new variables the action in
Eq. (\ref{sa}) assumes the form \begin{eqnarray}
 &  & {\cal S}\left[{\bf \Omega},\phi,{\varrho},\bar{h},h\right]={\cal S}_{B}[\bar{h},h]+\int_{0}^{\beta}d\tau{\cal H}_{{\bf \Omega,\phi}}[\rho,\bar{h},h]\nonumber \\
 &  & +{\cal S}_{0}\left[\phi\right]+2\sum_{{\bf r}}\int_{0}^{\beta}d\tau{\bf A}({\bf r}\tau)\cdot{\bf S}_{h}({\bf r}\tau),\label{sa2}\end{eqnarray}
 where ${\bf S}_{h}({\bf r}\tau)=\frac{1}{2}\sum_{\alpha\gamma}\bar{h}_{\alpha}({\bf r}\tau)\hat{\bm\sigma}_{\alpha\gamma}h_{\gamma}({\bf r}\tau)$.
Furthermore, \begin{eqnarray}
S_{0}[\phi]=\sum_{{\bf r}}\int_{0}^{\beta}d\tau\left[\frac{\dot{\phi}^{2}({\bf r}\tau)}{U}+\frac{1}{i}\frac{2\mu}{U}\dot{\phi}({\bf r}\tau)\right]\label{sphi}\end{eqnarray}
 stands for the kinetic and Berry term of the U(1) phase field in
the charge sector. The SU(2) gauge transformation in Eq. (\ref{sing2})
and the fermionic Berry term in Eq. (\ref{cberry}) generate SU(2)
potentials given by \begin{eqnarray}
{\bf R}^{\dagger}({\bf r}\tau){\partial}_{\tau}{\bf R}({\bf r}\tau) & = & {\bf R}^{\dagger}\left(\dot{\varphi}\frac{\partial}{\partial\varphi}+\dot{\vartheta}\frac{\partial}{\partial\vartheta}+\dot{\chi}\frac{\partial}{\partial\chi}\right){\bf R}\nonumber \\
 & = & -{\hat{\bm\sigma}}\cdot{\bf A}({\bf r}\tau),\end{eqnarray}
 where \begin{eqnarray}
A^{x}({\bf r}\tau) & = & \frac{i}{2}\dot{\vartheta}({\bf r}\tau)\sin\chi({\bf r}\tau)\nonumber \\
 & - & \frac{i}{2}\dot{\varphi}({\bf r}\tau)\sin\theta({\bf r}\tau)\cos\chi({\bf r}\tau)\nonumber \\
A^{y}({\bf r}\tau) & = & \frac{i}{2}\dot{\vartheta}({\bf r}\tau)\cos\chi({\bf r}\tau)\nonumber \\
 & + & \frac{i}{2}\dot{\varphi}({\bf r}\tau)\sin\theta({\bf r}\tau)\sin\chi({\bf r}\tau)\nonumber \\
A^{z}({\bf r}\tau) & = & \frac{i}{2}\dot{\varphi}({\bf r}\tau)\cos\vartheta({\bf r}\tau)+\frac{i}{2}\dot{\chi}({\bf r}\tau)\end{eqnarray}
 are the explicit expression for the vector potential in terms of
the Euler angles. 

\subsection{Integration over $V_{0}$ and $\varrho$}

We observe that the spatial and temporal fluctuations of the fields
$V_{0}({\bf r})$ and $\varrho({\bf r}\tau)$ will be energetically
penalized, since they are gapped and decouple from the angular and
phase variables. Therefore, in order to make further progress we subject
the functional to a saddle point HF analysis: the expectation value
of the static (zero frequency) part of the fluctuating electrochemical
potential $V_{0}(r)$ we calculate by the saddle point method to give
\begin{eqnarray}
V_{0}(r)=i\left(\mu-\frac{U}{2}n\right)\equiv i\bar{\mu}\end{eqnarray}
 where $n=\sum_{\alpha}\langle\bar{h}_{\alpha}({\bf r}\tau){h}_{\alpha}({\bf r}\tau)\rangle$
and the saddle point with respect to $\rho$ gives\begin{eqnarray}
\rho({\bf r}\tau)&=&(-1)^{{\bf r}}\Delta_{c}\nonumber \\
\Delta_{c}&=&U\langle S^{z}({\bf r}\tau)\rangle \label{spaff}
\end{eqnarray}
with $\Delta_c$ setting the magnitude
for the Mott-charge gap. The choice delineated in Eq. (\ref{spaff})
corresponds to the saddle point of the antiferomagnetic (with staggering
$\Delta_{c}$) type. Note, that the notion {}``antiferomagnetic\char`\"{}
here does not mean an actual long--range ordering - for this the angular
spin-quantization variables have to be ordered as well. The fermionic
sector, in turn, is governed by the effective Hamiltonian \begin{eqnarray}
 &  & {\cal H}_{{\bf \Omega,\phi}}=\sum_{{\bf r}}{\varrho}({\bf r}\tau)[\bar{h}_{{\uparrow}}({\bf r}\tau)h_{\uparrow}({\bf r}\tau)-\bar{h}_{{\downarrow}}({\bf r}\tau)h_{\downarrow}({\bf r}\tau)]\nonumber \\
 &  & -t\sum_{\langle{\bf r},{\bf r}'\rangle}\sum_{\alpha\gamma}\left[{\mathcal{U}}^{\dagger}({\bf r}\tau){\mathcal{U}}_{}({\bf r'}\tau)\right]_{\alpha\gamma}\bar{h}_{{\alpha}}({\bf r}\tau)h_{\gamma}({\bf r}'\tau)\nonumber \\
 &  & -\bar{\mu}\sum_{{\bf r}\alpha}\bar{h}_{\alpha}({\bf r}\tau)h_{\alpha}({\bf r}\tau),\label{explicit}\end{eqnarray}
 where $\bar{\mu}=\mu-nU/2$ is the chemical potential with a Hartree
shift originating from the saddle-point value of the static variable
$V_{0}$. 

\section{Spin-angular action}

Since we are interested in the magnetic properties of the system a
natural step is to obtain the effective action that involves the spin-directional
degrees of freedom ${\bf {\Omega}}$, which important fluctuations
correspond to rotations. This can be done by integrating out fermions:
\begin{eqnarray}
 &  & {\cal Z}=\int[{\cal D}\phi{\cal D}{\bf \Omega}]\int\left[{\cal D}\bar{h}{\cal D}{h}\right]e^{-{\cal S}[{\varphi,\phi,\vartheta},\bar{h},h]}\nonumber \\
 &  & \equiv\int\left[{\cal D}{\bf \Omega}\right]e^{-{\cal S}[{\bf \Omega}]},\label{explicit3}\end{eqnarray}
 where \begin{eqnarray}
{\cal S}[{\bf \Omega}]=-\ln\int[{\cal D}\phi{\cal D}\bar{h}{\cal D}{h}]e^{-{\cal S}[{\varphi,\phi,\vartheta},\bar{h},h]}\end{eqnarray}
 generates the cumulant expansions for the low energy action in the
form ${\cal S}[{\bf \Omega}]={\cal S}_{B}[{\bf \Omega}]+{\cal S}_{J}[{\bf \Omega}]$.

\subsection{Topological theta term}

In general, in addition to the usual exchange term, the action describing
antiferromagnetic spin systems is expected to have a topological Berry
phase term \begin{eqnarray}
{\cal S}_{B}[{\bf \Omega}] & = & -2\sum_{{\bf rr'}}\int_{0}^{\beta}d\tau{\bf A}({\bf r}\tau)\cdot\langle{\bf S}_{h}({\bf r'}\tau')\rangle,\end{eqnarray}
 where \begin{eqnarray}
 &  & \langle S_{h}^{z}({\bf r}\tau)\rangle=\frac{1}{2}(n_\uparrow -n_\downarrow)=\frac{\Delta_{c}}{U},\label{deltacdef}\end{eqnarray}
which results from the saddle point value in Eq. (\ref{spaff}).
 In terms of angular variables, the Berry term becomes \begin{equation}
{\cal S}_{B}[{\bf \Omega}]=\frac{\theta}{i}\sum_{{\bf r}}\int_{0}^{\beta}d\tau\left[\dot{\varphi}({\bf r}\tau)\cos\vartheta({\bf r}\tau)+\dot{\chi}({\bf r}\tau)\right].\label{sberry}\end{equation}
 If we work in Dirac {}``north pole\char`\"{} gauge ${\chi}({\bf r}\tau)=-{\varphi}({\bf r}\tau)$
one recovers the familiar form ${\cal S}_{B}[{\bf \Omega}]=\frac{\theta}{i}\sum_{{\bf r}}\int_{0}^{\beta}d\tau\dot{\varphi}({\bf r}\tau)[1-\cos\vartheta({\bf r}\tau)]$.
Here, the integral of the first term in Eq. ({\ref{sberry}) has
a simple geometrical interpretation as it is equal to a solid angle
swept by a unit vector ${\bf \Omega}(\vartheta,\varphi)$ during its
motion. The extra phase factor coming from the Berry phase, requires
some little extra care, since it will induce quantum mechanical phase
interference between configurations. In regard to the non-perturbative
effects, we realized the presence of an additional parameter with
the topological angle or so-called theta term \begin{equation}
\theta=\frac{{\Delta_{c}}}{U}\label{topoltheta}\end{equation}
 that is related to the Mott gap. In the large-$U$ limit one has
$\Delta_{c}\to U/2$, so that $\theta\to\frac{1}{2}$ relevant for
the half-integer spin. However, for arbitrary $U$ the theta term
will be different from that value, which, as we show will be instrumental
for destruction of the antiferomagnetic order away from the spin-localized
$U\to\infty$ limit. 

\subsection{AF exchange}

Now we proceed with the calculation of the exchange term in the spin-angular
action. We concentrate on the second order cumulant term in the hopping
element $t$ containing four fermion operators: \begin{widetext}
\begin{eqnarray}
{\cal S}^{(2)}[{\bar{h}},h]= &  & -\frac{t^{2}}{2}\int_{0}^{\beta}d\tau d\tau'\left\langle \sum_{|{\bf r}_{1}-{\bf r}_{1}'|=n.n.}\bar{z}({\bf r}_{1}\tau)z({\bf r}_{1}'\tau)\sum_{\alpha\alpha'}\left[{\bf R}^{\dagger}({\bf r}_{1}\tau){\bf R}_{}({\bf r'}_{1}\tau)\right]_{\alpha\alpha'}\bar{h}_{{\alpha}}({\bf r}_{1}\tau)h_{\alpha'}({\bf r}_{1}'\tau)\right.\times\nonumber \\
 &  & \times\left.\sum_{|{\bf r}_{2}-{\bf r}_{2}'|=n.n.}\bar{z}({\bf r}_{2}\tau')z({\bf r}_{2}'\tau')\sum_{\gamma\gamma'}\left[{\bf R}^{\dagger}({\bf r}_{2}\tau'){\bf R}({\bf r}_{2}'\tau')\right]_{\gamma\gamma'}\bar{h}_{{\gamma}}({\bf r}_{2}\tau')h_{\gamma'}({\bf r}_{2}'\tau')\right\rangle ,\end{eqnarray}
 \end{widetext} where $\langle\dots\rangle$ denotes averaging over
U(1) phase fields and and fermions. The averaging in the charge sector
is performed with the use of the U(1) phase action in Eq. (\ref{sphi})
to give \begin{eqnarray}
 &  & \langle\bar{z}({\bf r}_{1}\tau)z({\bf r}_{1}'\tau)\bar{z}({\bf r}_{2}\tau')z({\bf r}_{2}'\tau')\rangle\nonumber \\
 &  & \simeq(\delta_{{\bf r_{1},r_{1}'}}\delta_{{\bf r_{2},r_{2}'}}+\delta_{{\bf r_{1},r_{2}'}}\delta_{{\bf r'_{1},r_{2}}})\times\nonumber \\
 &  & \times\exp\left\{ -\frac{U}{2}\left[|\tau-\tau'|-\frac{(\tau-\tau')^{2}}{\beta}\right]\right\} .\label{chargecharge}\end{eqnarray}
 Furthermore, with the help of the gradient expansion \begin{eqnarray}
{\bf R}_{}({\bf r}\tau') & = & {\bf R}_{}({\bf r}\tau)+(\tau'-\tau)\partial_{\tau}{\bf R}_{}({\bf r}\tau)\nonumber \\
 & + & O[(\tau'-\tau)^{2}]\nonumber \\
{\bf h}_{}({\bf r}\tau') & = & {\bf h}_{}({\bf r}\tau)+(\tau'-\tau)\partial_{\tau}{\bf h}_{}({\bf r}\tau)\nonumber \\
 & + & O[(\tau'-\tau)^{2}]\end{eqnarray}
 we write the relevant part of the action in the form \begin{eqnarray}
 &  & {\cal S}_{J}[{\bf \Omega}]=-\frac{t^{2}}{2}\int_{0}^{\beta}d\tau d\tau'\exp\left[-\frac{U}{2}|\tau-\tau'|\right]\nonumber \\
 &  & \times\sum_{|{\bf r}-{\bf r}'|=a}\sum_{{\alpha\alpha'\atop \gamma\gamma'}}\left[{\bf R}^{\dagger}({\bf r}\tau){\bf R}_{}({\bf r'}\tau)\right]_{\alpha\alpha'}\left[{\bf R}^{\dagger}({\bf r}'\tau){\bf R}({\bf r}\tau)\right]_{\gamma\gamma'}\nonumber \\
 &  & \times\langle\bar{h}_{{\alpha}}({\bf r}\tau)h_{\gamma'}({\bf r}\tau)\rangle\langle h_{\alpha'}({\bf r}'\tau)\bar{h}_{{\gamma}}({\bf r}'\tau)\rangle.\end{eqnarray}
 In the low temperature limit (on the energy scale given by $U$),
by making use of the formula \begin{eqnarray}
\lim_{\tau\to0}\int_{0}^{\beta}d\tau'{e^{-\frac{{\left|\tau-\tau'\right|U}}{{2}}}}=\frac{{2}}{{U}}-\frac{{2e^{-\frac{{\beta U}}{{2}}}}}{{U}}\label{limcharge1}\end{eqnarray}
 and with the aid of the fermionic occupation numbers \begin{eqnarray}
 &  & \langle h_{\alpha'}({\bf r}'\tau)\bar{h}_{{\gamma}}({\bf r}'\tau)\rangle=(1-n_{\gamma})\delta_{\alpha'\gamma}\nonumber \\
 &  & \langle\bar{h}_{{\alpha}}({\bf r}\tau)h_{\gamma'}({\bf r}\tau')\rangle=n_{\alpha}\delta_{\alpha\gamma'}\end{eqnarray}
 we arrive at \begin{eqnarray}
{\cal S}_{J}[{\bf \Omega}]= &  & \frac{t^{2}}{U}\int_{0}^{\beta}d\tau\sum_{|{\bf r}-{\bf r}'|=a}\sum_{\alpha\gamma}\left[{\bf R}^{\dagger}({\bf r}\tau){\bf R}({\bf r'}\tau)\right]_{\alpha\gamma}\nonumber \\
 &  & \times\left[{\bf R}^{\dagger}({\bf r}'\tau){\bf R}({\bf r}\tau)\right]_{\gamma\alpha}(n_{\gamma}-1)n_{\alpha}.\end{eqnarray}
 Finally, making use of the following composition formula for the
SU(2) matrices \begin{eqnarray}
 &  & \left[{\bf R}^{\dagger}({\bf r}\tau){\bf R}({\bf r'}\tau)\right]_{\alpha\gamma}\left[{\bf R}^{\dagger}({\bf r'}\tau){\bf R}({\bf r}\tau)\right]_{\gamma\alpha}\nonumber \\
 &  & =\frac{1}{2}[1-{\bf \Omega}({\bf r}\tau)\cdot{\bf \Omega}({\bf r'}\tau)](1-\delta_{\alpha\gamma})\nonumber \\
 &  & +\frac{1}{2}[1+{\bf \Omega}({\bf r}\tau)\cdot{\bf \Omega}({\bf r'}\tau)]\delta_{\alpha\gamma}\end{eqnarray}
 we obtain the desired part of the spin action \begin{eqnarray}
{\cal S}_{J}[{\bf \Omega}] & = & \frac{J}{4}\sum_{\langle{\bf r}{\bf r}'\rangle}\int_{0}^{\beta}d\tau\left[{\bf \Omega}({\bf r}\tau)\cdot{\bf \Omega}({\bf r'}\tau)\right.\nonumber \\
 & + & \left.n(n-2)\right]\label{eq:SJ}\end{eqnarray}
 with the AF-exchange coefficient \begin{eqnarray}
 &  & J(\Delta_{c})=\frac{4t^{2}}{U}(n_{\uparrow}-n_{\downarrow})^{2}\equiv\frac{4t^{2}}{U}\left(\frac{2\Delta_{c}}{U}\right)^{2}.\label{afex}\end{eqnarray}
The factor $\sim t^2/U$ comes from integration over $U(1)$ charge degrees of freedom
(see, Eqs. (\ref{chargecharge}), (\ref{limcharge1})), whereas the occupation numbers result
from integration over fermionic variables (see, Eq. \ref{deltacdef}). 
From the Eq. (\ref{afex}) it is evident that for $U\to\infty$ one
has $J(\Delta_{c})\sim\frac{4t^{2}}{U}$ since $\frac{2\Delta_{c}}{U}\to1$
in this limit. In general the AF-exchange parameter persists as long
as the charge gap $\Delta_{c}$ exists. However, $J(\Delta_{c})$
diminishes rapidly in the $U/t\to0$ weak coupling limit, see Fig.
\ref{fig1}. 
%
\begin{figure}
\begin{centering}
\includegraphics[width=7.5cm]{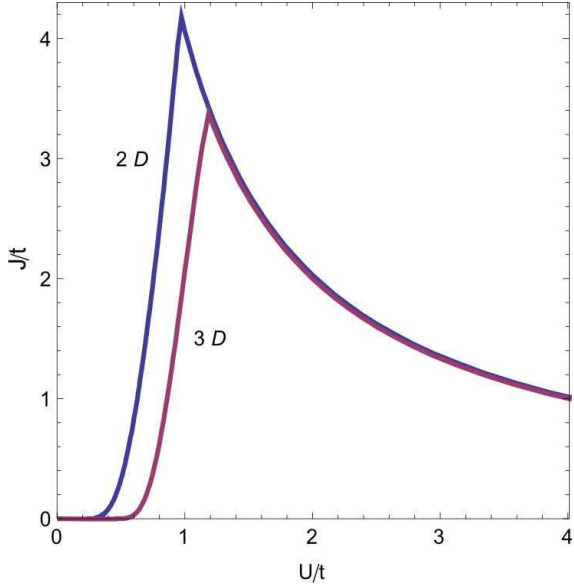} 
\par\end{centering}

\caption{(Color online) The antiferomagnetic exchange parameter $J$ as a
function of the Coulomb interaction $U$ in two and three dimensions.
The fermionic occupation number is fixed by $n=1$. }

\label{fig1} 
\end{figure}


\section{Fermionic sector}

Now we evaluate the effective interaction between fermions by integrating
out by means of cumulant expansion the gauge degrees of freedom. To this end we write the partition
function as \begin{eqnarray}
{\cal Z} & = & \int[{\cal D}\phi{\cal D}{\bf \Omega}]\int\left[{\cal D}\bar{h}{\cal D}{h}\right]e^{-{\cal S}[{\varphi,\phi,\vartheta},\bar{h},h]}\nonumber \\
 & \equiv & \int\left[{\cal D}\bar{h}{\cal D}{h}\right]e^{-{\cal S}[\bar{h},h]},\label{explicit1}\end{eqnarray}
 where \begin{eqnarray}
{\cal S}[\bar{h},h]=-\ln\int[{\cal D}\phi{\cal D}{\bf \Omega}]e^{-{\cal S}[{\varphi,\phi,\vartheta},\bar{h},h]}\end{eqnarray}
 generates cumulant expansion for the effective fermionic action.
Since, both U(1) and SU(2) gauge fields couple to the hopping element $t$, in 
the lowest order of cumulant expansion we reveal the hopping renormalization
$t\to tg$, \begin{eqnarray}
 &  & g=g_{c}g_{s}=\nonumber \\
 &  & g_{c}=\langle{\bar{z}}({\bf r}\tau){z}({\bf r'}\tau)\rangle\nonumber \\
 &  & g_{s}=\langle\left[{\bf R}^{\dagger}({\bf r}\tau){\bf R}_{}({\bf r'}\tau)\right]_{\uparrow\uparrow}\rangle\nonumber \\
 &  & =\langle[{\bf R}^{\dagger}({\bf r}\tau){\bf R}_{}({\bf r'}\tau)]_{\downarrow\downarrow}\rangle\label{ggcgs}\end{eqnarray}
 where the coefficients $g_{c}$ and $g_{s}$ contribute to the band
renormalization in a way, which is similar to the action of the Gutzwiller
factors,\cite{guc} and have to be calculated self consistently, according to the Eq. (\ref{ggcgs})
that involve charge and spin-angular correlation functions.
However, as long as there is no ordering in the charge sector, 
$\langle{\bar{z}}({\bf r}\tau){z}({\bf r'}\tau)\rangle=0$ and $g=0$ resulting in renormalized 
hopping $t=0$. Thus, we have to rest on the second order 
of the cumulant expansion, in which one obtains
a contribution to the effective action in the form \begin{equation}
{\cal S}^{(2)}[\bar{h},h]=-\frac{2t^{2}}{U}\int_{0}^{\beta}d\tau\sum_{\langle{\bf r}{\bf r}'\rangle}{\cal F}^{\dagger}({\bf r}\tau{\bf r}'\tau){\cal F}({\bf r}\tau{\bf r}'\tau)\end{equation}
 with the bond operators \begin{equation}
{\cal F}({\bf r}\tau{\bf r}'\tau)=\frac{{\bar{h}}_{\uparrow}({\bf r}\tau)h_{\uparrow}({\bf r'}\tau)+{\bar{h}}_{\downarrow}({\bf r}\tau)h_{\downarrow}({\bf r'}\tau)}{\sqrt{2}}\end{equation}
 Since the ${\cal S}^{(2)}[\bar{h},h]$ is quartic in the fermionic
variables, we resort to the Hubbard-Stratonovich decoupling with the
help of the complex variables defined on the links of the lattice
\begin{equation}
e^{{-{\cal S}}^{(2)}[\bar{h},h]}=\int[{\cal D}^{2}\xi]e^{-\sum\limits _{\langle{\bf r}{\bf r}'\rangle}\int\limits _{0}^{\beta}d\tau\left(\frac{2}{J}|\xi|^{2}+\xi{\cal \bar{F}}+\bar{\xi}{\cal F}\right)},\end{equation}
 where ${\cal D}^{2}\xi=\prod_{\langle{\bf r}{\bf r'}\rangle\tau}d^{2}\xi({\bf r}\tau{\bf r'}\tau)$
and $d^{2}\xi=d{\rm Re}\xi d{\rm Im}\xi$. Saddle point with respect
to $\xi$ gives \begin{equation}
\xi=\frac{J}{2}\langle{\cal F}({\bf r}\tau{\bf r}'\tau)\rangle=\frac{J}{2\sqrt{2}}\sum_{\alpha}\langle{\bar{h}}_{\alpha}({\bf r}\tau)h_{\alpha}({\bf r'}\tau)\rangle.\end{equation}
 Denoting \begin{equation}
v=\sum_{\alpha}\langle{\bar{h}}_{\alpha}({\bf r}\tau)h_{\alpha}({\bf r'}\tau)\rangle,\end{equation}
 which plays the role of the kinetic energy parameter for fermions
we obtain \begin{equation}
{\cal S}^{(2)}[\bar{h},h]=-\frac{Jv^{2}}{4}+t_{J}\sum_{\langle{\bf rr'}\rangle\alpha}[{\bar{h}}_{\alpha}({\bf r}\tau)h_{\alpha}({\bf r'}\tau)+h.c.],\end{equation}
%
\begin{figure}
\begin{centering}
\includegraphics[width=7.5cm]{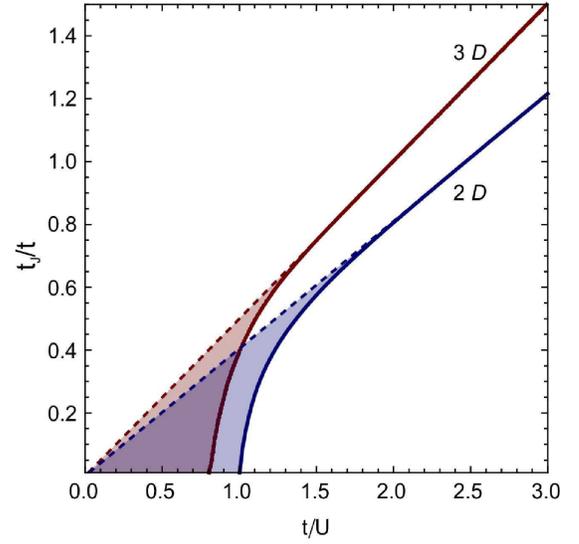} 
\par\end{centering}

\caption{(Color online) The effective hopping parameter $t_{J}/t$ (see, Eq. (\ref{effectivehoping})) related to
the AF-exchange as a function of $t/U$ at $n=1$ for two and three
dimensional Hubbard model at $n=1$.}

\label{fig2} 
\end{figure}
where 
\begin{equation}t_{J}=\frac{1}{4}Jv\label{effectivehoping}\end{equation} is the effective hopping parameter that involves
the antiferromagnetic exchange parameter $J$. The interaction dependence
of this quantity is depicted in Fig. \ref{fig2}. Summarizing the
results we obtain for the fermionic action \begin{eqnarray}
{\cal S}[\bar{h},h]={\cal S}_{B}[\bar{h},h]+\int_{0}^{\beta}d\tau{\cal H}[\bar{h},h],\end{eqnarray}
 with the effective Hamiltonian \begin{eqnarray}
{\cal H}[\bar{h},h] &  & =\sum_{{\bf r}}(-1)^{{\bf r}}\Delta_{c}[\bar{h}_{{\uparrow}}({\bf r}\tau)h_{\uparrow}({\bf r}\tau)-\bar{h}_{{\downarrow}}({\bf r}\tau)h_{\downarrow}({\bf r}\tau)]\nonumber \\
 &  & -(tg-t_{J})\sum_{\langle{\bf r},{\bf r}',\alpha\rangle}\bar{h}_{{\alpha}}({\bf r}\tau)h_{\alpha}({\bf r}'\tau)\nonumber \\
 &  & -\bar{\mu}\sum_{{\bf r}\alpha}\bar{h}_{\alpha}({\bf r}\tau)h_{\alpha}({\bf r}\tau).\label{explicit4}\end{eqnarray}
The result of the gauge transformations is that we have managed to
cast the strongly correlated problem into a system of mutually non-interacting
fermions, submerged in the bath of strongly fluctuating U(1) and SU(2)
fields, whose dynamics is governed by the energy scale set by the
Coulomb interaction $U$ coupled to fermions via hopping term and
with the Zeeman-type contribution with the massive field ${\varrho}({\bf r}\tau)$
related to the Mott gap $\Delta_{c}$. To calculate the latter one
has to introduce two inequivalent sublattices, let say A and B and
write the Hamiltonian in terms of two sublattice operators in the
reduced Brillouin zone (RBZ). The fermionic propagator then reads
\begin{eqnarray}
{\bf G}({\bf k}\omega_{n})=\frac{\left(\begin{array}{cc}
i\nu_{n}+\epsilon_{{\bf k}}-\bar{\mu} & \Delta_{c}\\
\Delta_{c} & i\nu_{n}-\epsilon_{{\bf k}}-\bar{\mu}\end{array}\right)}{(i\nu_{n}-\epsilon_{{\bf k}}-\bar{\mu})(i\nu_{n}+\epsilon_{{\bf k}}+\bar{\mu})-\Delta_{c}^{2}},\end{eqnarray}
 where $\nu_{n}=\pi(2n+1)/\beta$, $n=\pm1,\pm2,\dots$ are the fermionic
Matsubara frequencies. The self-consistency equations for the Mott gap,
the kinetic energy bond parameter and fermionic occupation number
are given by \begin{eqnarray}
\Delta_{c} & = & \frac{1}{\beta N}{\sum_{{\bf k}\nu_{n},\alpha}}'{G}_{\uparrow\downarrow}({\bf k}\nu_{n})\nonumber \\
v & = & \frac{1}{\beta N}{\sum_{{\bf k}\nu_{n},\alpha}}'\gamma({\bf k}){G}_{\alpha\alpha}({\bf k}\nu_{n})\nonumber \\
n & = & 1-\frac{1}{\beta N}{\sum_{{\bf k}\nu_{n},\alpha}}'{G}_{\alpha\alpha}({\bf k}\nu_{n}),\end{eqnarray}
 where the sums with prime index denote the summations over wave vectors
inside the the RBZ and $\gamma({\bf k})=(\cos k_{x}+\cos k_{y})/2$.
By performing the summations over Matsubara frequencies one obtains
explicitly with the use of the fermionic distribution $n_{F}(x)=1/(e^{\beta x}+1)$:
\begin{eqnarray}
1=\frac{U}{2N}\sum_{{\bf k}}\frac{{n_{F}\left(-{E_{{\bf k}}}-\bar{\mu}\right)}-{n_{F}\left({E_{{\bf k}}}-\bar{\mu}\right)}}{E_{{\bf k}}}\label{gapsol}\end{eqnarray}
 for the gap parameter, \begin{eqnarray}
v=\frac{1}{2N}\sum_{{\bf k}}\gamma({\bf k})\frac{{n_{F}\left(-{E_{{\bf k}}}-\bar{\mu}\right)}-{n_{F}\left({E_{{\bf k}}}-\bar{\mu}\right)}}{E_{{\bf k}}},\end{eqnarray}
 the fermion kinetic energy parameter, \begin{eqnarray}
n=\frac{1}{N}\sum_{{\bf k}}\left[{n_{F}\left(-{E_{{\bf k}}}-\bar{\mu}\right)}+{n_{F}\left({E_{{\bf k}}}-\bar{\mu}\right)}\right]\end{eqnarray}
 and the occupation number, respectively. At half-filling, solutions
of the Eq. (\ref{gapsol}) for the gap $\Delta_{c}$ are stabilized
for any arbitrarily small $U$, however by computing of the free energy
one can find that no stable antiferromagnetic solutions away from
$n=1$ exists.\cite{stab} Solutions of the self-consistency equation
for the gap $\Delta_{c}$ leading to densities away from half-filling
correspond to maxima instead of minima in the free energy. If a certain
occupation near the half-filling is enforced, the system will experience
a phase separation in a half-filled antiferromagnetic and a non-half-filled
paramagnetic region. %

\section{CP$^{1}$ formulation}

Since the fermionic field can be systematically integrated out, the
main practical difficulty comes from the dynamics of spin-directional
fluctuations. To proceed with the spin-bosonic action we resort to
the CP$^{1}$ representation (see, e.g. Ref.\onlinecite{auer}).
In the CP$^{1}$ representation, the SU(2) rotation matrix is expressed
in terms of two Schwinger bosons, \begin{eqnarray}
{\bf R}({\bf r}\tau)=\left[\begin{array}{cc}
\zeta_{1}({\bf r}\tau) & -\bar{\zeta}_{2}({\bf r}\tau)\\
\zeta_{2}({\bf r}\tau) & \bar{\zeta}_{1}({\bf r}\tau)\end{array}\right]\end{eqnarray}
 with the constraint $|\zeta_{1}({\bf r}\tau)|^{2}+|\zeta_{2}({\bf r}\tau)|^{2}=1$.
The unimodular constraint can be resolved by using the Euler angles
parametrization \begin{eqnarray}
\zeta_{1}({\bf r}\tau) & = & e^{-\frac{i}{2}[\varphi({\bf r}\tau)+\chi({\bf r}\tau)]}\cos\left[\frac{\vartheta({\bf r}\tau)}{2}\right]\nonumber \\
\zeta_{2}({\bf r}\tau) & = & e^{\frac{i}{2}[\varphi({\bf r}\tau)-\chi({\bf r}\tau)]}\sin\left[\frac{\vartheta({\bf r}\tau)}{2}\right],\label{cp1}\end{eqnarray}
 which make link between the $\zeta_{1}({\bf r}\tau)$, $\zeta_{2}({\bf r}\tau)$
fields and ${\bf {\Omega}({\bf r}\tau)}$ variables. By definition
\begin{equation}
{\bf S}_{\zeta}({\bf r}\tau)=\frac{1}{2}\sum_{\alpha\gamma}\bar{\zeta}_{\alpha}({\bf r}\tau)\hat{\bm\sigma}_{\alpha\gamma}\zeta_{\gamma}({\bf r}\tau)\equiv\frac{1}{2}{\bf \Omega}({\bf r}\tau)\end{equation}
 are the {}``bosonic\char`\"{} spins in the complex-projective (CP$^{1}$)
formulation, while the action, see Eq. (\ref{eq:SJ}) becomes\begin{eqnarray}
{\cal S}_{J}[{\bf \Omega}] & \to & J\sum_{\langle{\bf r}{\bf r}'\rangle}\int_{0}^{\beta}d\tau\left[{\bf S}_{\zeta}({\bf r}\tau)\cdot{\bf S}_{\zeta}({\bf r'}\tau)-\frac{1}{4}\right].\label{afaction}\end{eqnarray}
 Consequently, the complete spin-bosonic action ${\cal S}[{\bar{\zeta}},{\zeta}]={\cal S}_{\theta}[{\bar{\zeta}},{\zeta}]+{\cal S}_{J}[{\bar{\bm\zeta}},{\bm\zeta}]$
reads \begin{eqnarray}
{\cal S}_{\theta}[{\bar{\zeta}},{\zeta}] & = & -2\theta\sum_{{\bf r}\alpha}(-1)^{{\bf r}}\int_{0}^{\beta}d\tau\bar{\zeta}_{\alpha}({\bf r}\tau)\dot{\zeta}_{\alpha}({\bf r}\tau)\nonumber \\
{\cal S}_{J}[{\bar{\zeta}},{\zeta}] & = & -J\sum_{\langle{\bf r}{\bf r}'\rangle}\int_{0}^{\beta}d\tau\bar{{\cal A}}({\bf r}\tau{\bf r'}\tau){\cal A}({\bf r}\tau{\bf r'}\tau),\end{eqnarray}
 where $\theta={\frac{\Delta_{c}}{U}}$ is the {}``theta angle\char`\"{}
parameter in the Berry-phase term, while the AF-exchange term ${\cal S}_{J}[{\bar{\zeta}},{\zeta}]$
we write with the help of the valence-bond operators ${\cal A}({\bf r}\tau{\bf r'}\tau)$
for which the following relations hold \begin{eqnarray}
 &  & {\bf S}_{\zeta}({\bf r}\tau)\cdot{\bf S}_{\zeta}({\bf r'}\tau)=-\bar{{\cal A}}({\bf r}\tau{\bf r'}\tau){\cal A}({\bf r}\tau{\bf r'}\tau)+\frac{1}{4}\nonumber \\
 &  & {\cal A}({\bf r}\tau{\bf r'}\tau)=\frac{{\zeta}_{\uparrow}({\bf r}\tau){\zeta}_{\downarrow}({\bf r}'\tau)-{\zeta}_{\downarrow}({\bf r}\tau){\zeta}_{\uparrow}({\bf r}'\tau)}{\sqrt{2}}.\end{eqnarray}

\subsection{HS decoupling}

In order to achieve a consistent representation of the underlying
antiferromagnetic structure, it is unavoidable to explicitly split
the degrees of freedom according to their location on sublattice A
or B. Since the lattice is bipartite allowing one to make the unitary
transformation \begin{eqnarray}
{\zeta}_{\uparrow}({\bf r}\tau) & \to & -{\zeta}_{\downarrow}({\bf r}\tau)\nonumber \\
{\zeta}_{\downarrow}({\bf r}\tau) & \to & {\zeta}_{\uparrow}({\bf r}\tau)\end{eqnarray}
 for sites on one sublattice, so that %
\begin{eqnarray}
{\cal A}({\bf r}\tau{\bf r'}\tau) & \to & {\cal A}'({\bf r}\tau{\bf r'}\tau)=\sum_{\alpha=1}^{2}\frac{{\zeta}_{\alpha}({\bf r}\tau){\zeta}_{\alpha}({\bf r}'\tau)}{\sqrt{2}}.\end{eqnarray}
 Biquadratic (four-variable) terms in the Lagrangian cannot be readily
integrated in the path integral. Introducing a complex variable for
each bond that depends on {}``imaginary time\char`\"{} $Q({\bf r}\tau{\bf r'}\tau)$
we decouple the four-variable terms $\bar{{\cal A}}'({\bf r}\tau{\bf r'}\tau){\cal A}'({\bf r}\tau{\bf r'}\tau)$
using the formula \begin{equation}
e^{{\cal S}_{J}[{\bar{\bm\zeta}},{\bm\zeta}]}=\int[{\cal D}^{2}Q]e^{-\sum\limits _{\langle{\bf r}{\bf r}'\rangle}\int\limits _{0}^{\beta}d\tau\left(\frac{2}{J}|Q|^{2}+Q{\bar{\bm\zeta}}\cdot\bar{\bm\zeta}+\bar{Q}{\bm\zeta}\cdot{\bm\zeta}\right)}\end{equation}
 where ${\cal D}^{2}Q=\prod_{\langle{\bf r}{\bf r'}\rangle\tau}d^{2}Q({\bf r}\tau{\bf r'}\tau)$
and $d^{2}Q=d{\rm Re}Qd{\rm Im}Q$. To handle the unimodularity condition
one introduces Lagrange multipliers $\lambda({\bf r}\tau)$ at each
time and site. Then with the help of the Dirac-delta functional \begin{equation}
\delta\left(|{\bm\zeta}({\bf r}\tau)|^{2}-1\right)=\int\left[\frac{{\cal D}\lambda}{2\pi i}\right]e^{\sum\limits _{{\bf r}}\int\limits _{0}^{\beta}d\tau\lambda(|{\bm\zeta}|^{2}-1)}\end{equation}
the variables $\zeta_{1}({\bf r}\tau)$,$\zeta_{2}({\bf r}\tau)$
are now unconstrained bosonic fields. Thus, the local constraints
are reintroduced into the theory through the dynamical fluctuations
of the auxiliary $\lambda$ field \begin{equation}
{\cal Z}=\int[{\cal D}^{2}Q{\cal D}^{2}{\bm\zeta}{\cal D}\lambda]e^{-\sum\limits _{\langle{\bf r}{\bf r}'\rangle}\int\limits _{0}^{\beta}d\tau\left(\frac{2|Q|^{2}}{J}-\lambda\delta_{{\bf r}{\bf r}'}+{\cal H}_{Q}[\bar{\bm\zeta},{\bm\zeta}]\right)},\end{equation}
 where \begin{equation}
{\cal H}_{Q}[\bar{\bm\zeta},{\bm\zeta}]=\sum_{\langle{\bf r}{\bf r}'\rangle}\int_{0}^{\beta}d\tau\left(Q{\bar{\bm\zeta}}\cdot\bar{\bm\zeta}+\bar{Q}{\bm\zeta}\cdot{\bm\zeta}+\lambda{\bar{\bm\zeta}}\cdot{\bm\zeta}\right).\end{equation}
 Furthermore, one then performs a saddle-point approximation over
the $Q$ and $\lambda$ fields \begin{eqnarray}
Q_{{\rm sp}}({\bf r}\tau{\bf r}'\tau) & = & -\frac{J}{2}\sum_{\alpha=1}^{2}\langle\bar{\zeta}_{\alpha}({\bf r}\tau)\bar{\zeta}_{\alpha}({\bf r}'\tau)\rangle\nonumber \\
 & = & -\frac{J}{\sqrt{2}}\langle{\cal A}'({\bf r}\tau{\bf r'}\tau)\rangle\nonumber \\
\frac{1}{2} & = & \frac{1}{2}\sum_{\alpha=1}^{2}\langle\bar{\zeta}_{\alpha}({\bf r}\tau){\zeta}_{\alpha}({\bf r}\tau)\rangle\end{eqnarray}
 by assuming the uniform solution $Q_{{\rm sp}}({\bf r}\tau{\bf r}'\tau)\equiv Q$
we obtain for the Hamiltonian in the spin-bosonic sector \begin{eqnarray}
{\cal H}_{Q}[\bar{\bm\zeta},{\bm\zeta}] & = & \frac{1}{\beta N}\sum_{{\bf k}\omega_{n}}\sum_{\alpha=1}^{2}[\bar{\zeta}_{{\alpha}}({\bf k},\omega_{n}),{\zeta}_{{\alpha}}(-{\bf k},-\omega_{n})]\nonumber \\
 & \times & \frac{\hat{{\cal G}}_{\alpha}^{-1}({\bf k},\omega_{n})}{2}\left[\begin{array}{c}
{\zeta}_{{\alpha}}({\bf k},\omega_{n})\\
\bar{\zeta}_{{\alpha}}(-{\bf k},-\omega_{n})\end{array}\right]\end{eqnarray}
 with \begin{equation}
\hat{{\cal G}}_{\alpha}^{-1}({\bf k},\omega_{n})=\left(\begin{array}{cc}
2i\theta\omega_{n}+{\lambda} & -z\gamma_{{\bf k}}Q\\
-z\gamma_{{\bf k}}Q & -2i\theta\omega_{n}+{\lambda}\end{array}\right).\end{equation}
 Subsequently, performing the sums over Matsubara frequencies one
obtains \begin{eqnarray}
Q & = & \frac{J(\Delta_{c})}{N}\sum_{{\bf k}}\frac{1}{2\theta}\frac{z\gamma_{{\bf k}}^{2}Q}{2\omega_{{\bf k}}}\coth\left(\frac{\beta\omega_{{\bf k}}}{4\theta}\right)\nonumber \\
1 & = & -\frac{1}{2\theta}+\frac{1}{N}\sum_{{\bf k}}\frac{1}{2\theta}\frac{\lambda}{\omega_{{\bf k}}}\coth\left(\frac{\beta\omega_{{\bf k}}}{4\theta}\right),\label{selfconsistentordereqs}\end{eqnarray}
 where $\omega_{{\bf k}}=\sqrt{\lambda^{2}-(z\gamma_{{\bf k}}Q)^{2}}$
and $z$ is the lattice coordination number. 

\section{AF long-range order parameter}

A characteristic property of strongly correlated systems is the existence
of local moments. Weak-coupling theories usually fail to properly
describe these local moments. When electron correlation effects become
stronger, in general, spin fluctuations have to be considered seriously.
The HF transition temperature bears a physical meaning as a temperature
below which the amplitude $\Delta_{c}$ of the AF order parameter
takes a well-defined value. This is also interpreted as the appearance
of local moments. However, a nonzero value of $\Delta_{c}$ does not
imply the existence of AF long--range order. For this the angular
degrees of freedom ${\bf \Omega}({\bf r}\tau)$ have also to be ordered,
whose low-lying excitations are in the form of spin waves. In the
CP$^{1}$ representation (where the Neel field is represented by two
Schwinger bosons) Bose-Einstein condensation of the Schwinger bosons
at zero temperature signals the appearance of AF long-range order.
The AF order parameter in terms of the original fermion operators
is defined as \begin{eqnarray}
m_{AF} & = & \sum_{{\bf r}}(-1)^{{\bf r}}\langle S^{z}({\bf r}\tau)\rangle=\nonumber \\
 & = & \sum_{{\bf r}}(-1)^{{\bf r}}\langle{\bf \Omega}({\bf r}\tau)\rangle\cdot\langle{\bf S}_{h}({\bf r}\tau)\rangle.\end{eqnarray}
 Owing the fact that $\langle S_{h}^{a}({\bf r}\tau)\rangle=(-1)^{{\bf r}}\Delta_{c}\delta_{a,z}$
we obtain \begin{eqnarray}
m_{AF} & = & \Delta_{c}\sum_{{\bf r}}\langle{\Omega}^{z}({\bf r}\tau)\rangle\nonumber \\
 & = & \Delta_{c}\sum_{{\bf r}}\left[\langle{\bar{\zeta}}_{\uparrow}({\bf r}\tau){\zeta}_{\uparrow}({\bf r}\tau)\rangle-\langle{\bar{\zeta}}_{\downarrow}({\bf r}\tau){\zeta}_{\downarrow}({\bf r}\tau)\rangle\right].\end{eqnarray}
 Furthermore, the order parameter for the CP$^{1}$ {}``boson condensate\char`\"{}
is \begin{eqnarray}
\langle{\bar{\zeta}}_{\alpha}({\bf k}\omega_{n})\rangle & = & \langle{\zeta}_{\alpha}({\bf k}\omega_{n})\rangle\nonumber \\
 & = & \sqrt{\frac{\beta N}{2}}m_{0}\delta_{0,\omega_{n}}\delta_{\uparrow,\alpha}\left(\delta_{{\bf k},0}+\delta_{{\bf k},{\bf Q}}\right).\end{eqnarray}
 This yields a macroscopic contribution (i.e., order one) to the staggered
magnetization and represents a macroscopic contribution to the CP$^{1}$
bosons density, of the $\alpha=\uparrow$ bosons at the mode with
${\bf k}=0,\omega_{n}=0$ thus giving \begin{eqnarray}
m_{AF} & = & \frac{\Delta_{c}}{\beta UN}\sum_{{\bf k},\omega_{n}}\left[\langle{\bar{\zeta}}_{\uparrow}({\bf k}\omega_{n}){\zeta}_{\uparrow}({\bf k}\omega_{n})\rangle\right.\nonumber \\
 &  & \left.-\langle{\bar{\zeta}}_{\downarrow}({\bf k}\omega_{n}){\zeta}_{\downarrow}({\bf k}\omega_{n})\rangle\right]=\frac{\Delta_{c}}{U}m_{0}^{2}.\end{eqnarray}
 Finally, the fraction of condensed Schwinger bosons is given by \begin{eqnarray}
m_{0}^{2} & = & 1+\frac{1}{2\theta_s}-\frac{1}{N}\sum_{{\bf k}}\frac{1}{2\theta_{s}}\frac{\lambda}{\omega_{{\bf k}}}\coth\left(\frac{\beta\omega_{{\bf k}}}{4\theta_{s}}\right),\end{eqnarray}
 which represents the extension of the saddle point equation for the
Lagrange multiplier to the region of the ordered state. 

\subsection{$d=2$ Hubbard model }

In two dimensions, we expect no long-range AF order at finite temperatures
due to Mermin and Wagner's theorem. This could be verified by explicitly
performing two-dimensional momentum summations in Eqs. (\ref{selfconsistentordereqs})
with the help of density of states for the two-dimensional
square lattice $\rho_{2d}(\epsilon)=\int_{-\pi}^{\pi}[d^{2}{\bf k}/(2\pi)^{3}\delta[\epsilon-\epsilon({\bf k})]$,
where $\epsilon({\bf k})=\epsilon_{0}(\cos k_{x}+\cos k_{y})$ with
\begin{eqnarray}
\rho_{2d}(\epsilon) & = & \frac{1}{\pi^{2}\epsilon}\Theta\left(1-\frac{|{\epsilon}|}{2\epsilon_{0}}\right)\times\nonumber \\
 & \times & {{\bf K}\left[\sqrt{1-\left(\frac{\epsilon}{2\epsilon_{0}}\right)^{2}}\right]}.\end{eqnarray}
As a result, there is no antiferromagnetic order at finite temperature in two-dimensional
Hubbard model. Taking the zero temperature limit in Eq. (\ref{selfconsistentordereqs}) and 
fixing the fermionic occupation number at $n=1$, in the limit $U/t\to\infty$
we find the order parameter value, $m_{AF}=0.308$ in the ground state 
(in agreement with the calculations from the Ref. \onlinecite{delannoy})
that is less than the classical value $S=1/2$. Monte Carlo calculation
on $2D$ Hubbard model\cite{hirsh} gave $m_{AF}=0.4$ This effect
is due to the quantum zero-point motion, which has its origin in the
noncommutability of the Hamiltonian and the staggered magnetization.
In the opposite weak coupling limit, $U/t\to0$, the gap $\Delta_{c}$
persists at arbitrary small value of $U/t$, however the true order
parameter $m_{AF}$, which involves also the density of condensed
Schwinger bosons vanishes at $U/t\approx0.621$ (see, Fig. \ref{fig3}).
The destruction of the AF order is due to the Berry phase term $\theta$ whose
coefficient, cf. Eq. (\ref{topoltheta}) differs from the localized
spin value $S=1/2$ while entering the weak coupling limit. 
In particular, for $U/t\rightarrow 0$, that is in the weak coupling limit, 
$\theta$ goes to zero along with the charge gap $\Delta_c$ (see, Eq. (\ref{topoltheta})) and the self
consistency Eqs. (\ref{selfconsistentordereqs}) predict vanishing
of the long-range AF order.

%
\begin{figure}
\begin{centering}
\includegraphics[width=7.5cm]{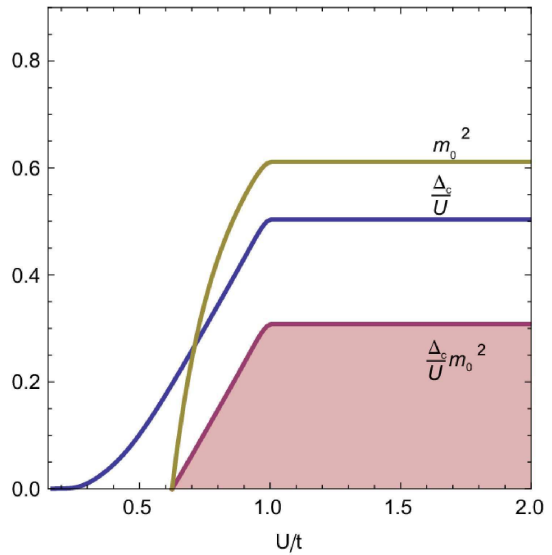} 
\par\end{centering}

\caption{(Color online) The Mott gap $\Delta_{c}$, fraction of condensed Schwinger
bosons $m_{0}^{2}$ and AF order parameter $m_{AF}$ for the half-filled
Hubbard two-dimensional model at zero temperature. }

\label{fig3} 
\end{figure}

%
\begin{figure}
\begin{centering}
\includegraphics[width=7.5cm]{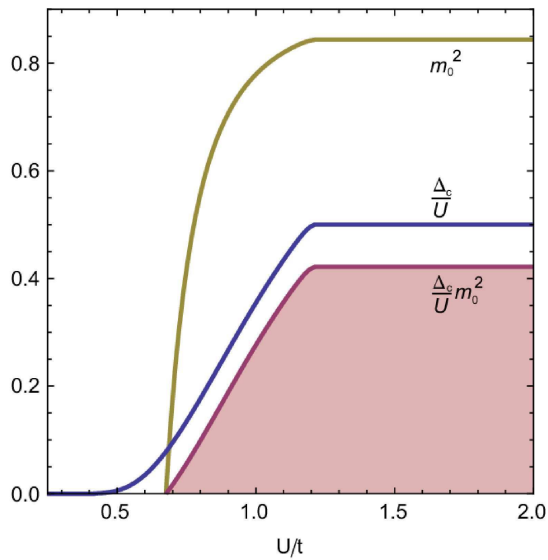} 
\par\end{centering}

\caption{(Color online) Same as in Fig. \ref{fig2} but for three-dimensional
Hubbard model. }

\label{fig4} 
\end{figure}


\subsection{$d=3$ Hubbard model }

In three dimensions, for a system with an ordered ground state, thermally
excited states reduce the spin correlations at finite temperatures.
When the temperature is much higher than the typical coupling energy
scale $J$, we expect the spins to be uncorrelated at large distances
and the magnetization $m_{AF}$ to vanish in the absence of an ordering
field. This requires a phase transition at some temperature $T_{c}$
between the ordered and disordered phases. As in the previous case
we employ the density of states for the cubic lattice $\rho_{3d}(\epsilon)=\int_{-\pi}^{\pi}[d^{3}{\bf k}/(2\pi)^{3}\delta[\epsilon-\epsilon({\bf k})]$,
where $\epsilon({\bf k})=\epsilon_{0}(\cos k_{x}+\cos k_{y}+\cos k_{z})$:
\begin{eqnarray}
\rho_{3d}(\epsilon) & = & \frac{1}{\pi^{3}\epsilon}\int_{a_{1}}^{a_{2}}dx\Theta\left(1-\frac{|{\epsilon}|}{3\epsilon_{0}}\right)\times\nonumber \\
 & \times & \frac{{\bf K}\left[\sqrt{1-\left(\frac{\epsilon}{2\epsilon_{0}}+\frac{x}{2}\right)^{2}}\right]}{\sqrt{1-x^{2}}}\end{eqnarray}
 with $a_{1}={\rm max}(-1,-2-\epsilon/\epsilon_{0})$, $a_{2}={\rm min}(1,2-\epsilon/\epsilon_{0})$.
%
\begin{figure}
\begin{centering}
\includegraphics[width=7.5cm]{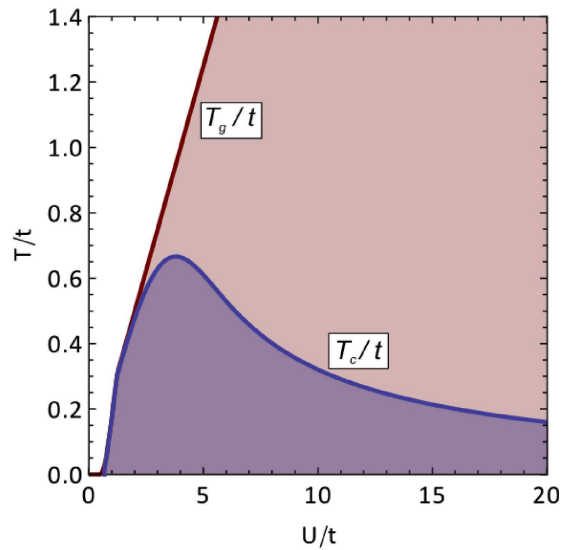} 
\par\end{centering}

\caption{(Color online) The temperature-interaction phase diagram for the three
dimensional Hubbard model at half filling. Depicted is the temperature
$T_{g}$ for the vanishing of the gap parameter $\Delta_{c}$ as well
as the true critical temperature $T_{c}$ at which the log-range AF
order ceases to exist, signalled by vanishing of $m_{AF}$. }

\label{fig5} 
\end{figure}

%
\begin{figure}
\begin{centering}
\includegraphics[width=7.5cm]{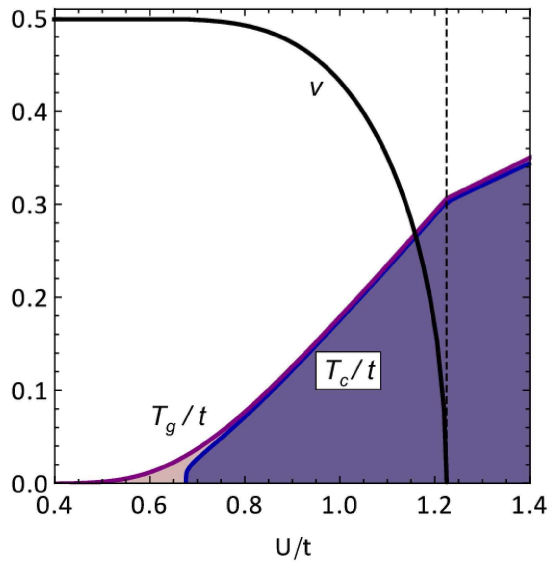} 
\par\end{centering}

\caption{(Color online) Details of the phase diagram from Fig. \ref{fig5}
in the weak-coupling regime. Depicted is also the interaction dependence
of the kinetic parameter $v$ that enters the effective, $J$-induced
hopping parameter $t_{J}$. Note the vanishing of the true AF order
for $U/t\approx0.676$ and persistence of the gap $\Delta_{c}$ for
arbitrary value of $U/t$. }

\label{fig6} 
\end{figure}

The interaction dependence of the AF magnetic moment is depicted in
Fig. \ref{fig4}. In the $U\to\infty$ localized limit it is $m_{AF}\approx0.422$,
i.e less the mean field value $\Delta_{c}/U=1/2$, however bigger
than in the case of the $2D$ Hubbard model. Finally, Fig. \ref{fig5}
displays the calculated antiferromagnetic phase diagram as a function
of temperature and interaction strength. At weak coupling our theory
clearly describes a Slater antiferromagnet with an exponentially
small AF gap. As $U$ increases, the Slater antiferromagnet progressively
evolves into a Mott-Heisenberg antiferromagnet with an AF gap of order
$U$. In the weak interaction limit there is a destruction of the
AF order at $U/t=0.676$ (see Fig. \ref{fig6}), due to the topological
Berry phase term whose coefficient, cf. Eq. (\ref{topoltheta}) deviates
from the from the localized spin value $S=1/2$ in the weak coupling
limit $U/t$. The AF critical temperature has a maximum at $U/t\approx3.78$.
It is worthwhile to compare our results with the previous work on
the subject. Numerical methods such as dynamical cluster approximation
(DCA)\cite{dca,maier} give $U/t\approx7.5$, whereas dynamical mean-field
theory approximation (DMFA)\cite{dmfa} predicts $U/t\approx10$.
The methods based on a perturbation theory with respect to the interaction
strength\cite{pt1,pt2} are unable to reproduce the maximum in the
AF critical temperature as a function of $U/t$. The significantly
higher values of $U/t$ resulting from DCA and DFMA have to be explained
by the restricted ability of these methods while handling spatial
fluctuations. Regarding the value of maximum of the critical temperature
$T_{c}/t\approx0.667$ found here, it agrees with the result of Monte
Carlo simulations by Scalettar et al\cite{mc} $T_{c}/t\approx0.72$
and by Hirsch\cite{mc2} who obtained $T_{c}\approx W/18t$, where
$W=12t$ is the bandwidth for the $3D$ Hubbard model, i.e. $T_{c}/t\approx0.666$.


\section{Conclusions}

In conclusion, we have investigated the ground state properties of
the two-dimensional half-filled one band Hubbard model in the strong
large-$U$ to intermediate coupling limit i.e., away from the strict
Heisenberg limit and antiferromagnetic phase diagram of the three
dimensional Hubbard model using SU(2)$\times$U(1) rotating reference
frame description. Our focus on systems in the strong to intermediate
coupling regime, was motivated by the fact that weaker interactions
are leading to increased electron mobility, which in turn should reduce
the stability of magnetic phases. Calculations with the Hamiltonian
for interacting electrons were reduced to calculation of functional
integrals with a phase-angular action. Collective bosonic fields are
introduced by means of a Hubbard-Stratonovich decoupling of the Hubbard
interaction and subsequent gauge transformation. Our implementation
for the Hubbard model is consistent with the spin rotation symmetry
and simultaneously is able to reproduce the Hartree-Fock result. One
important technical aspect arising in the construction of effective
theories is that electron-defined operators in the bare high-energy
theory are transformed into the composite particles subsequently employed
in calculations within the effective low-energy theory. The inclusion
of the quantum and spatial fluctuations has been shown to have a dramatic
effect on transition temperatures and phase diagram. We have also
compared the outcome of our calculations to a number of methods that
were employed by other authors. \acknowledgments One of us (T.K.K)
acknowledges the support by the Ministry of Education and Science
(MEN) under Grant No. 1P03B 103 30 in the years 2006-2008. T.A.Z.
wants to thank the Foundation for Polish Science for support within
{}``Grants for Scholars'' framework (12/06).

\end{document}